\documentstyle[osa,manuscript]{revtex}  % DON'T CHANGE

\begin{document}

\title{Airy-like patterns in heavy ion elastic scattering}

\author{R.Anni}
\address{
Dipartimento di Fisica dell'Universit\`a, Universit\`a di Lecce, Lecce, Italia\\
Istituto Nazionale di Fisica Nucleare, Sezione di Lecce, Lecce, Italia
}
\maketitle

\begin{abstract}
% DON'T CHANGE THIS LINE
%
A semiclassical analysis of an optical potential cross section is
presented. The cross section considered is characterized by the appearance
of an {\it Airy-like} pattern. This pattern is similar to that which is present in
many cross sections, which fit the recent measurements of light heavy ion
elastic scattering, and is considered as a manifestation of a rainbow
phenomenon.

The semiclassical analysis shows that, in the case considered, the
oscillations arise from the interference between the contributions from two
different terms of a multi-reflection expansion of the scattering function,
and, therefore, cannot be associated with the rainbow phenomenon.
\end{abstract}

%\twocolumn

The elastic differential cross sections of
$^{16}$O+$^{16}$O\cite{KHO00,NIC00} and $^{16}$O +$^{12}$C\cite{NIC99,OGL00},
recently measured at several energies and over wide angular ranges,
are characterized by the appearance of structures which are rather well
reproduced using optical potentials with deep real and shallow 
imaginary parts. A shallow imaginary part allows significant contributions 
from the internal region and this suggests that the gross structures in the angular
distribution can be explained as arising from the contributions 
of trajectories refracted by the deep real potential.

In order to isolate the contributions from these refracted trajectories,
the simple decomposition of the scattering amplitude in 
near- and far-side components\cite{FUL75} is commonly used: 
because, usually, for strongly absorbing potentials only the near-side 
component significantly contributes to the cross sections, it results 
rather natural to think that the far-side component should retain 
the contributions from trajectories penetrating the internal region.

Applying this decomposition  to the optical potential scattering
amplitude an {\it Airy-like} pattern often appears in the far-side
cross section and this has stimulated the claim that one is observing a {\it 
rainbow} phenomenon.

We remember that the meteorological rainbow phenomenon is produced by the
scattering of light by the water-droplets of rain and that a simple,
scalar, model of the process is provided by the non-relativistic scattering
by a spherical well.

The semiclassical limit for scattering by this potential was discussed in
detail by Nussenzveig\cite{NUS69a,NUS69b} in the framework of an exact
multi-reflection expansion of the scattering function, named Debye
expansion, in which the $n$-th term accounts for the contributions of
trajectories which are refracted $n-1$ times in the internal region.

In this multi-reflection expansion the primary rainbow is associated with
the third term, retaining the contribution from trajectories which
propagates two-times in the internal region. Mathematically, the rainbow
oscillations arise from the coherent superposition of the contributions from
two saddle-points, coalescing at the rainbow scattering angle. In a
neighborhood of this angle, the use of uniform asymptotic techniques allows
one to express the scattering amplitude in term of an Airy function whose
maximum replaces the singularity predicted by the non uniform method.

In order to confirm the rainbow nature of the spectacular {\it Airy-like} pattern
observed in some far-side cross sections it seems desirable to look
for the two saddle points contributions which, coalescing at the rainbow scattering 
angle, should produce the Airy maximum.

In the extreme semiclassical limit, these two saddle point
contributions should be obtained directly from the exact scattering function 
$S_l$. In this limit the derivative of the argument of $S_l$, with
respect to the angular momentum $l$, is just the classical deflection
function, which should show a maximum or a minimum at the rainbow angle. In
practical cases, the derivative of $\arg(S_l)$ presents a more or less
marked oscillatory behavior that prevents the treatment of this quantity as a
deflection function. Owing to this it is not possible to obtain the saddle point 
contributions by simply using the exact $S_l$.

The same happens also for the scattering from a spherical well, and the reason
is that the link between the scattering angle and the angular momentum must be
looked for in each term of the multi-reflection expansion and not in the
exact scattering function.

Unfortunately, at present time, there does not exist an exact multi-reflection 
expansion for scattering by a generic potential equivalent to the
Debye expansion for the spherical well potential. However, a non uniform 
semiclassical method was proposed\cite{KNO76} for potentials with an 
arbitrary number of turning points in the complex $r-$plane, and 
a uniform semiclassical technique was developed\cite{BRI77} for the cases
in which only three turning points give the main contribution. From both 
these methods one can derive approximate
multi-reflection expansions\cite{KNO76,ANN81} in which the different
terms have the same physical meaning as the corresponding
terms in the exact Debye expansion for the scattering by a spherical well.

In this brief note we present the results obtained by analyzing, with the
uniform multi-reflection expansion, the scattering by one optical potential
whose far-side cross section exhibits a striking {\it Airy-like} pattern. 
The undesired result obtained is that the {\it Airy-like} 
pattern does not arise from interference between two saddle points in the same term 
of the multi-reflection expansion, but from interference between a saddle point
from the second term of the expansion, describing trajectories refracted 
in the internal region, with a contribution from the first term of the expansion,
describing trajectories which do not penetrate the internal region.
This last contribution is responsible for the {\it Fraunhofer-like} pattern in the
cross section of the first term of the expansion, supporting the
conjecture that it must be considered a diffractive contribution.

The optical potential here considered is one of those obtained\cite{OGL00} by fitting the
elastic scattering cross section of  $^{16}$O + $^{12}$C at $E_{Lab}=132$
MeV. This potential has conventional Saxon-Woods form factors with parameters $
V_{0}=282.2$ MeV, $R_{v}=2.818$ fm and $d_{v}=.978$ fm, for the real part, and $
W_{0}=13.86$ MeV, $R_{w}=5.689$ fm and $d_{w}=.656$ fm, for the imaginary part.
The only modification introduced in the optical potential here used, with 
respect to the original one, is represented by the use for the Coulomb part 
of a proper analytical potential, in order to allow the continuation of the 
quantities needed in a semiclassical analysis outside the real $r$-axis. 
The differences to the cross section values produced by this substitution 
are completely irrelevant.

In Fig.1 we show the modulus of $S_l$ and a rough estimate of the 
derivative of its argument, obtained using the formula 
$\Theta(\lambda)=\arg(S_{l+1})- \arg (S_{l})$, for integer $l$ values and with
$\lambda = l+\frac{1}{2}$.
The oscillatory behavior of $\arg (S_{l})$, that in the following we name 
the quantum deflection function, in the range of $l$ values around $l \simeq 23$
cancels the hope of estimating the cross section by applying the saddle point 
technique starting from the exact $S_{l}$. 
However the smooth behavior of both $\Theta(\lambda)$ and $|S_l|$ for 
$l$ values up to about 18 can be considered a signature of the dominance of
a classical contribution in this $l$ range. 

This behavior of $\Theta(\lambda)$  is just the one expected 
for the deflection function of trajectories refracted in the internal 
region of an attractive potential, and should produce a saddle point  
contribution to the  far-side cross section. 

A rough estimate of this  contribution, from the integer $l$ 
values up to 18, is shown by the open dots in Fig. 2.  It is orders of 
magnitude smaller than the far-side cross section (medium thickness dashed curve)
at forward angles, but increases for increasing angles until it becomes of 
the same order of magnitude as the far-side cross section  in the region in 
which the oscillations become more marked.

The simplest quantity in which to look for another saddle point
contribution, which interfering with the above one could produce the
{\it Airy-like} pattern, is the scattering function of the naive WKB 
approximation in which the imaginary part of the potential is treated 
as a perturbation\cite{BRO74}. This quantity  has an argument 
whose derivative with respect to $\lambda$ coincides with the classical 
deflection function calculated with only the real part of the potential.
In the present case this deflection function has a minimum of about
-310.15 degrees, at $\lambda \simeq 23.56$, and indicates the existence of
the two desired saddle point contributions.

The contributions from the first of these two branches of
the deflection function (thin dashed line in Fig. 2) closely follows the open
dots, the second (thin continuous line) results larger at forward angles, 
but not enough to justify the average behavior of the exact far-side 
cross section. The interference between the amplitudes of the two contributions
was not calculated; in any case it is evident that their sum does not 
exhibit any classical rainbow singularity, a singularity that 
the uniform technique should transform in the Airy maximum. 
Because the minimum of the deflection function is of -310.15 degrees, 
this singularity is expected at an angle of 49.85 degrees in the 
contributions to the cross section from trajectories coming from the 
near-side of the scattering plane.

These difficulties simply reflect the fact that the naive WKB 
approximation is too rough for a quantitative analysis of the cross section, 
and this is confirmed by the comparison of the cross section that it 
predicts (dotted line in Fig. 2) with the exact one (heavy continuous line).

The reason of the failure of the naive WKB approximation must be looked for
in the fact that the addition of a small imaginary part to a real potential
can dramatically modify the motion of the turning points, as function of 
the angular momentum.

In Fig. 3 are shown the positions, for integer $l$ values, of the turning points
nearest to the real $r$ axis (open dots), and of the orbiting points at which
two turning points coalesce, for complex $l$ values in this case. The small dots 
refer to the complete potential and the large ones to its real part. The  
squares show the singularities of the potential nearest to the real axis.

The trajectory of the real turning point, for the real potential,
is broken into two branches for the complete one: the first terminating in a 
location near to a singularity of the real part, the second originating near
a singularity of the imaginary part. In the following the turning points of these two branches will be indicated with the subscript 3 and 1.

The trajectory described by the turning point, with 
positive imaginary part and ending in a position near a singularity of 
the real potential, is not qualitatively modified. In the following this
turning point will be referred with the subscript 2. 

The addition of the imaginary part modifies the old trajectory
with negative imaginary part. The new trajectory starts in a location near to the
old one but ends in a location near a singularity of the imaginary potential.

This turning point and the new one, appearing in the first quadrant
near a singularity of the imaginary potential, remain far from the real axis and 
their contributions will be neglected.

Retaining only the contributions from the turning points labeled from 1 to
3 the uniform semiclassical multi-reflection expansion of $S(\lambda)$ is given by: 
\begin{equation}
S_{SC}(\lambda)=\sum_{n=0}^\infty S_n (\lambda),
\end{equation}
where 
\begin{equation}
S_0(\lambda)= \frac{{\rm exp}(2i\delta_1)} {N(\frac {S_{21}} {\pi}) },
\end{equation}
and, for $n \ge 1$, 
\begin{equation}
S_{n}(\lambda)=-(-)^n \frac{{\rm exp}[2i(nS_{32}+S_{21}+\delta_1)]} {N^{n+1}(%
\frac {S_{21}} {\pi})}.
\end{equation}
In the above equations $\delta_1$ is the complex WKB phase shift for the 
turning point $r_1$, $S_{ij}$ is the action integral, in units of $\hbar$, 
between the turning points $r_i$ and $r_j$, and $N(z)$ is the {\it barrier 
penetrability factor} given by: 
\begin{equation}
N(z)=\frac{\sqrt{2\pi}}{\Gamma(\frac{1}{2}+z)}{\rm exp}(z {\rm ln} z -z).
\end{equation}
As in the Debye expansion, the first term, usually denominated the {\it 
barrier} term, retains the contributions from trajectories not penetrating
the internal region, while the $n$-th term retains the contributions from
trajectories refracted $n$ times in the internal region with $n-1$
reflections at the turning point $r_2$.

The modulus and the quantum deflection function of the first two terms of
the expansion are shown in Fig. 4. The modulus of the second term (thick dashed line) 
is much larger than that of the first (thick continuous line) for small $l$ values, 
but it decreases while the other increases, until they become equal at $l \simeq 21$. 
For higher $l$ value the modulus of the first term rapidly increases while that of the second even more rapidly decreases.

The quantum deflection function of the first term (medium continuous line) has the typical 
behavior, apart from a small neighborhood of the grazing angular momentum, of the deflection function of trajectories reflected at the surface 
of a spherical region, in presence of an external small Coulomb field.

The quantum deflection function of the second term (medium dashed line) closely follows
the classical deflection function (dotted line) for $l$-values smaller than about 20; 
for higher values of $l$ it shows a rainbow behavior, less deep and more large 
than that predicted by classical mechanics. The very 
small value of the  modulus of the scattering function suggests that the
saddle point contribution from the branch of the quantum deflection function to the 
right of the rainbow angular momentum should result completely negligible.  

The thin curves represent the cubic spline interpolations, of the integer $l$ 
values, of the modulus and the quantum deflection function of the sum of the first 
two terms of the expansion. These curves are in very good agreement 
with the dots representing the corresponding exact quantities and provide a 
simple explanation of the origin of the irregular behavior of the 
exact quantum deflection function, as arising from the interference between the
contributions of two simpler component scattering functions.

The semiclassical cross section, and its near- and far-side components, obtained using in the partial wave expansion
the first two terms of the multi-reflection expansion, are shown in panel
{\it (a)} of Fig. 5. The differences between these quantities and the  corresponding
exact ones cannot be appreciated within the scale and the thickness used for
the curves. 

In panel {\it (b)} the complete 
semiclassical cross section is shown together with the separate cross sections of the 
first two terms of the expansion. The cross section of 
the first term is characterized by the appearance, at 
forward angles, of a {\it Fraunhofer-like} oscillatory pattern, while no 
{\it Airy-like} oscillatory pattern is present in that of the second term.
The {\it Airy-like} oscillatory pattern appears only in the complete cross
section and arises from the interference between the scattering amplitudes of
the two terms.

The near-  and far-side 
decompositions of the cross section of
the first and the second term of the expansion are shown in panels
{\it (c)} and {\it (d)}, respectively.
With the exclusion of the extreme backward scattering angles the cross
section of the second term of the expansion is completely far-side
and results in very good agreement with the dots that represent the
saddle point contribution to the cross section previously estimated using
the exact scattering function.

The {\it Fraunhofer-like} oscillatory behavior of the first term of the expansion
arises from the interference between a near-side and a far-side
contribution. It is just the interference between the far-side contributions 
of the first two terms which
is responsible for the {\it Airy-like} pattern appearing in the far-side 
component of the complete cross section.

Previous analyses of similar decompositions\cite{ANN81} have shown that 
the far-side component of the {\it barrier} term of the expansion 
retains the contribution from generalized diffracted 
trajectories. In the uniform semiclassical approximation for the {\it barrier}
scatterimg function, this contribution should, mathematically,  derive from 
the Sommerfeld pole (or if one likes: the barrier top resonance\cite{FRI77})
located near to the real $\lambda $-axis, at the $\lambda _{0}$ value for 
which $S_{21}(\lambda _{0})=-\frac{\pi }{2}$.

The correctness of this interpretation can only be proved by the direct
numerical calculation of the location and of the residue of
this pole. In any case, the analysis of the cross section here considered shows 
that the oscillations in the far-side cross section arise from the 
interference of contributions from different terms of the multi-reflection 
expansion. From this it follows that (if one 
agrees to reserve the {\it rainbow} denomination to the phenomena having the same 
justification of the phenomenon observed in meteorology) the use of the rainbow 
terminology for these oscillations should be avoided.

Irrespective of the nature of the other contribution, and on the denomination
of the interference pattern, the present analysis confirms that one of these two contributions is  a saddle point one, and that it is associated with trajectories
which more or less deeply penetrate the internal region and give important 
contributions to the optical potential cross section.

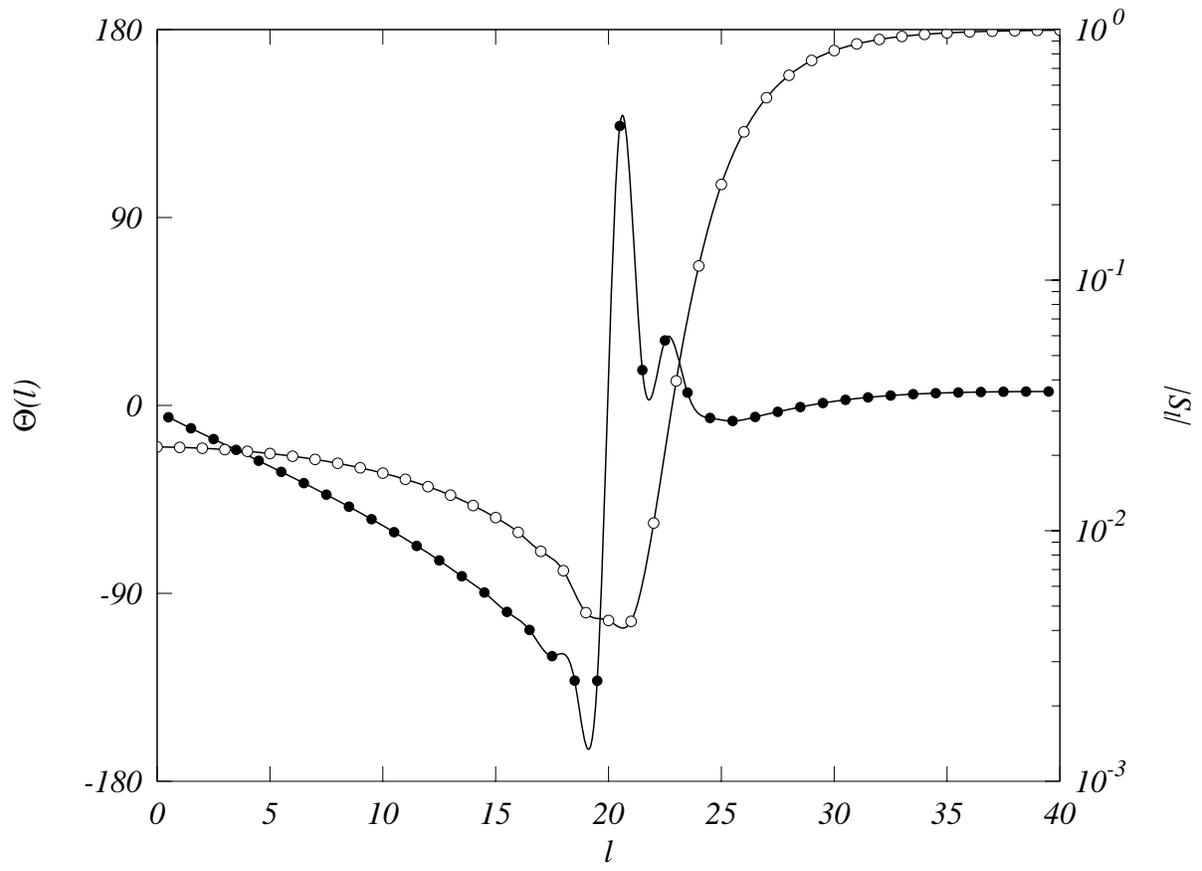
\begin{figure}
\caption{Modulus (open dots) and derivative of the argument (full dots) of the scattering function. The thin curves show the cubic spline interpolation of the dots.}
\end {figure}

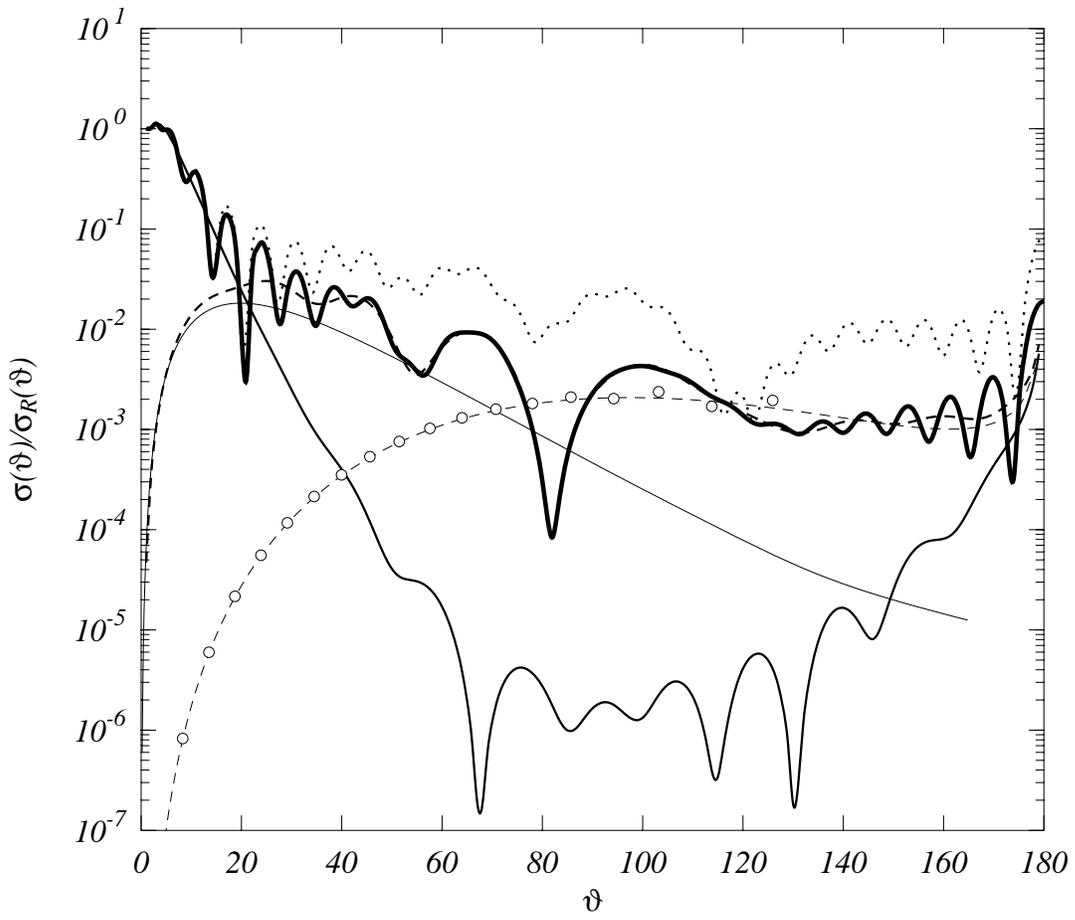
\begin{figure}
\caption{Cross section (heavy thick line), near- and far-side components (medium continuous and dashed lines), together with the naive WKB cross section
(dotted line) and the saddle point contribution from the two far-side branches of the WKB deflection function (thin continuous and dashed lines). The dots show the saddle point contribution estimated using the exact scattering function.}
\end {figure}

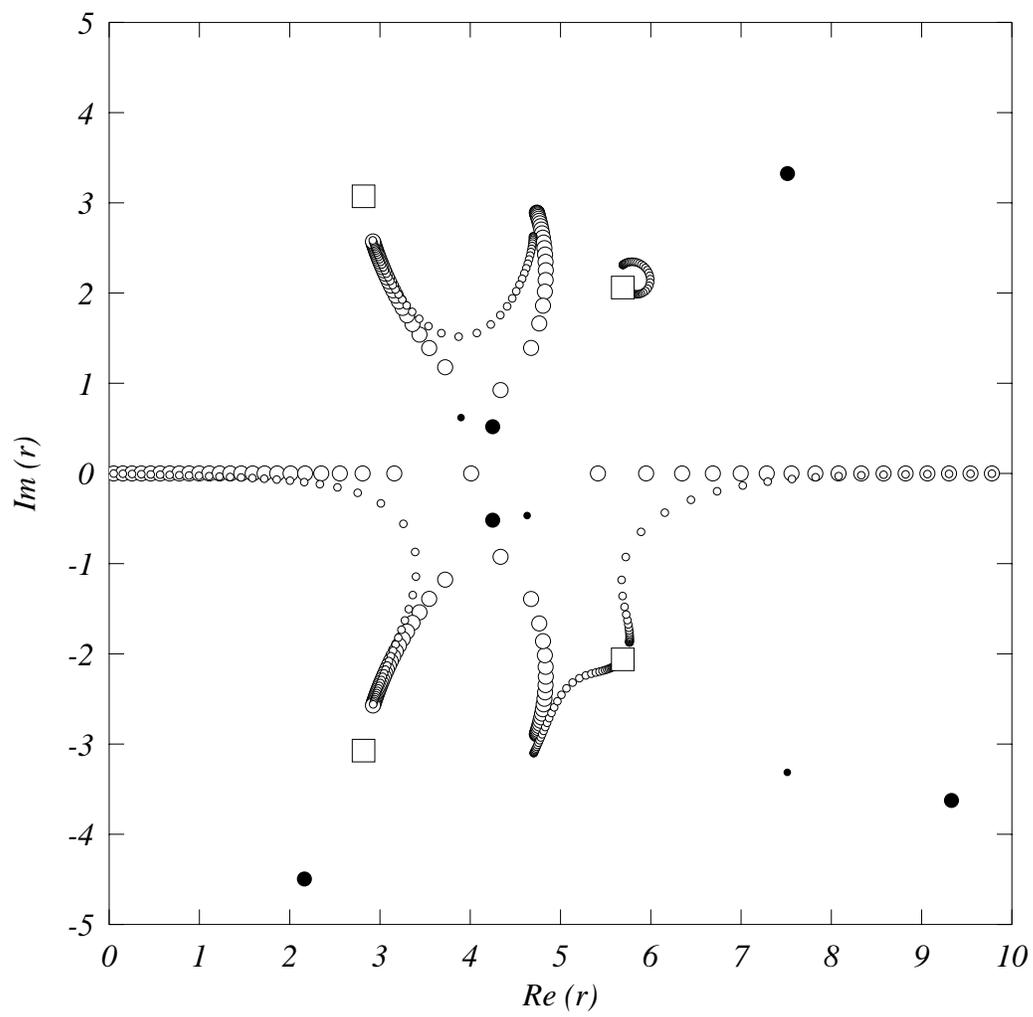
\begin{figure}
\caption{Turning points in the complex $r$-plane (open dots) and orbiting points
(full dots) for the complete potential (small) and for only the real part (large).
The squares indicate the singularities of the potential.}
\end {figure}

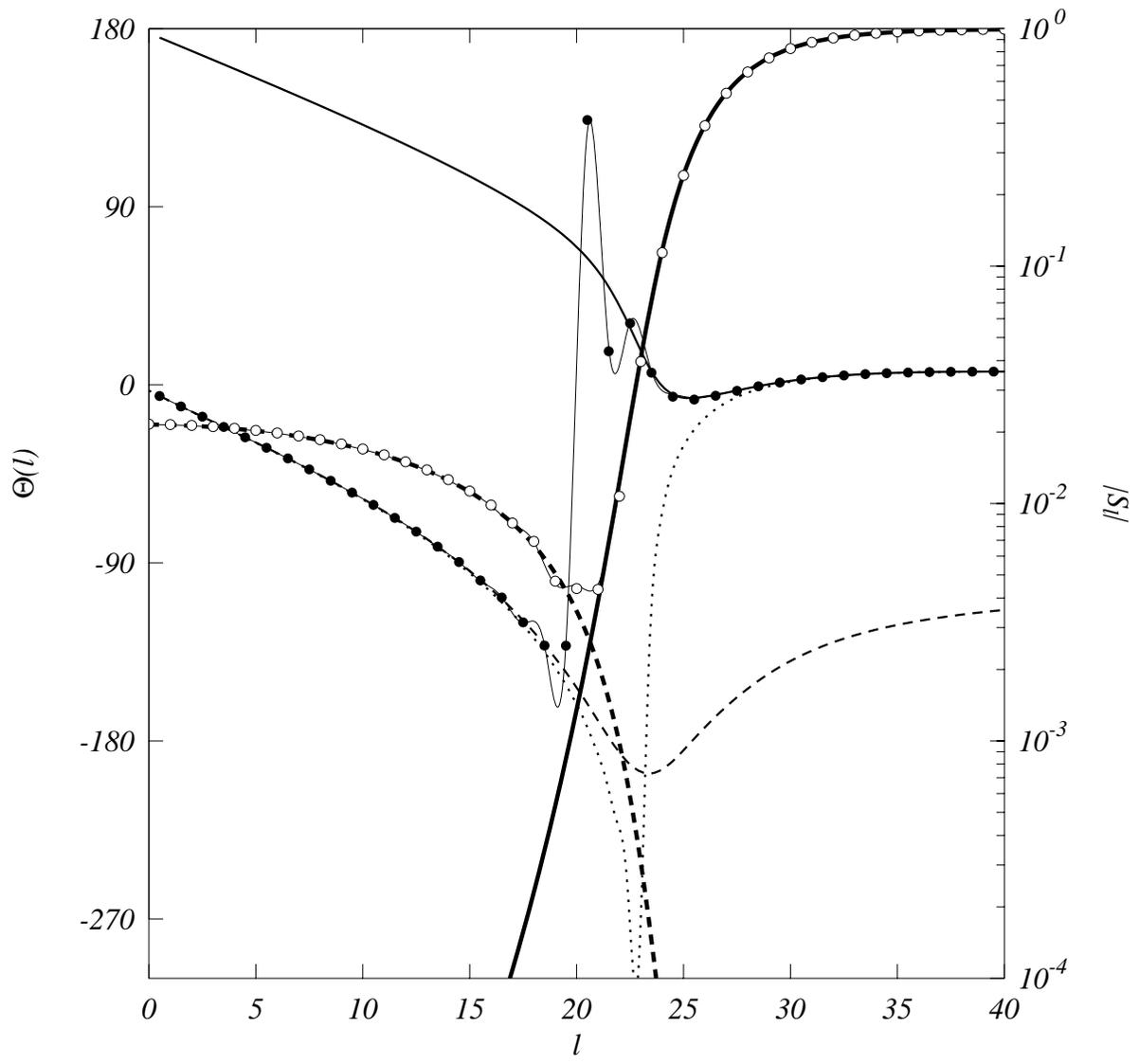
\begin{figure}
\caption{Modulus and deflection function (heavy and medium thickness lines)
of the first and second terms (continuous and dashed lines) of the multi-reflection
expansion. The thin lines show the same quantities for the sum of the two terms. The
the dots are from Fig. 1, and the dotted line shows the classical deflection function of the real potential.}

\end {figure}

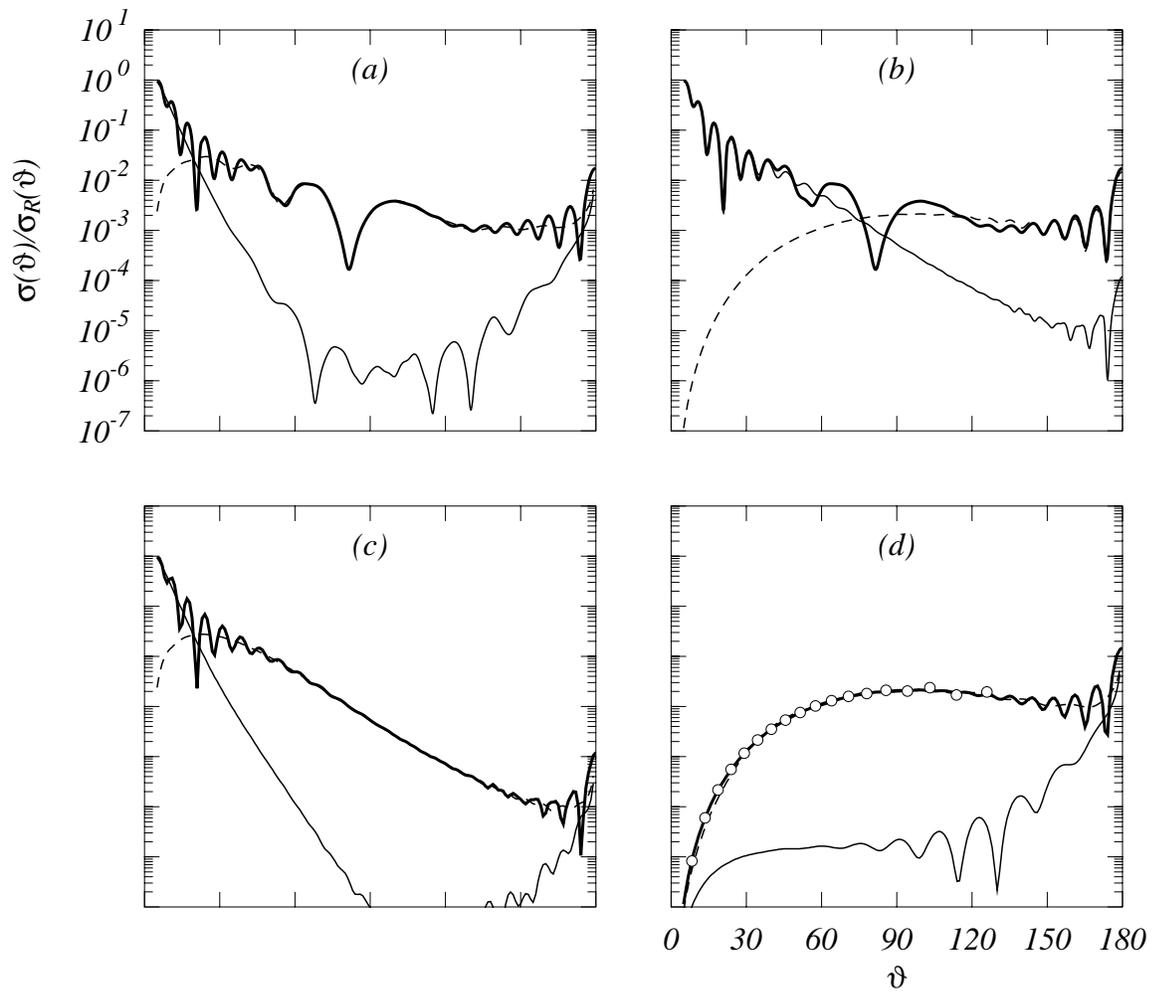
\begin{figure}
\caption{
(a) near- and far-side decomposition (continuous and dashed lines) of the 
semiclassical cross section (thick line); 
(b) cross section  of the first and second terms (continuous and dashed lines) of the multi-reflection expansion and of their sum  (thick line);
(c)  near- and far-side  decomposition (continuous and dashed lines)
of the  cross section of the first term of the expansion (thick line);
(d) the same as (c) for the second term of the expansion.
}

\end {figure}


\begin{references}
%% Please use the \bibitem command to create references.
% \bibitem{RefTag}Author, "Title," Journal {\bf Volume,}
% page numbers (year).  %(Format for Journal Reference)

\bibitem{KHO00}  Dao T. Khoa, W. von Oertzen, H. G. Bohlen, and F. Nuoffer, 
Nucl. Phys. {\bf A672} 387 (2000).

\bibitem{NIC00}  M. P. Nicoli, F. Haas, S. Szilner, Z. Basrak, A. Morsad, 
G. R. Satchler, and M. E. Brandan,  Phys. Rev. C {\bf 61} 034609 (2000).

\bibitem{NIC99}  M. P. Nicoli, F. Haas, R. M. Freeman, N. Aissaoui, C. Beck, 
A. Elanique, R. Nouicer, S. Szilner, Z. Basrak, A. Morsad,  M. E. Brandan, and
G. R. Satchler,  Phys. Rev. C {\bf 60} 064608 (2000).

\bibitem{OGL00}  A. A. Ogloblin, Yu. A. Glukhov, W. H. Trzaska, A. S. Dem'yanova, S. A. Goncharov, R. Julin, S. V. Klebnikov, M. Mutterer, M. V.Rozhkov, V. P. Rudakov, G. P. Tiorin, Dao T. Khoa, and G. R. Satchler,  Phys. Rev. C {\bf 62} 044601 (2000).

\bibitem{FUL75}  R. C. Fuller,  Phys. Rev. C {\bf 12} 1561 (1975).

\bibitem{NUS69a}  H. M. Nussenzweig,  J. Math. Phys. {\bf 10} 82 (1969).

\bibitem{NUS69b}  H. M. Nussenzweig,  J. Math. Phys. {\bf 10} 125 (1969).

\bibitem{KNO76}  J. Knoll and R. Schaeffer,  Ann. Phys. (N.Y.) {\bf 97} 307
(1976).

\bibitem{BRI77}  D. Brink and N. Takigawa,  Nucl. Phys. {\bf A279} 159 (1977).

\bibitem{ANN81}  R. Anni and L. Renna,  Nuovo Cimento A {\bf 42} 311 (1981).

\bibitem{BRO74}  R. Broglia, S. Landowne, R. A. Malfliet, V. Rostokin and Aa. Winther, 
Phys. Rep. C {\bf 11} 1 (1974).

\bibitem{FRI77}  W. A. Friedman and C .J. Goebel,  Ann. Phys. (N.Y.) {\bf 104} 145
(1977). 

\end{references}
\end{document}